%% file: wh.tex
\documentclass[]{elsarticle}
\usepackage{lineno}
\usepackage{hyperref}
\usepackage{cleveref}
\usepackage{natbib}
\usepackage{booktabs}  
\usepackage{graphicx}

\modulolinenumbers[5]
\journal{Energy and Buildings}

\begin{document}

\begin{frontmatter}

\title{Domain Knowledge Aids in Signal Disaggregation; the Example of the Cumulative Water Heater}

\author[hw,uc]{Alexander Belikov\fnref{eqc}}

\author[hw]{Guillaume Matheron\corref{mycorrespondingauthor}\fnref{eqc}}
\cortext[mycorrespondingauthor]{Corresponding author}
\ead[url]{www.hellowatt.fr}
\ead{guillaume_pub@matheron.eu}

\author[hw]{Johan Sassi\fnref{eqc}}

\address[hw]{Hello Watt, 48 rue René Clair 75018 Paris, France}
\address[uc]{Knowledge Lab, University of Chicago, Chicago, IL, USA}
\fntext[eqc]{Equal contribution. The source dataset is available at \href{https://osf.io/ebjd3/?view_only=54c6097e95984b6c9c933a38e574a8f7}{OSF}.}

\begin{abstract}
In this article we present an unsupervised low-frequency method aimed at detecting and disaggregating the power used by Cumulative Water Heaters (CWH) in residential homes.
Our model circumvents the inherent difficulty of unsupervised signal disaggregation by using both the shape of a power spike and its time of occurrence to identify the contribution of CWH reliably.
Indeed, many CHWs in France are configured to turn on automatically during off-peak hours only, and we are able to use this domain knowledge to aid peak identification despite the low sampling frequency.

In order to test our model, we equipped a home with sensors to record the ground-truth consumption of a water heater.
We then apply the model to a larger dataset of energy consumption of Hello Watt users consisting of one month of consumption data for 5k homes at 30-minute resolution.
In this dataset we successfully identified CWHs in the majority of cases where consumers declared using them.
The remaining part is likely due to possible misconfiguration of CWHs, since triggering them during off-peak hours requires specific wiring in the electrical panel of the house.
Our model, despite its simplicity, offers promising applications: detection of mis-configured CWHs on off-peak contracts and slow performance degradation.
\end{abstract}


\end{frontmatter}

\section{Introduction}\label{sec:introduction}


\paragraph{Smart meters in France}
In an effort to record and optimize energy consumption of households as well as facilitate billing, starting from the beginning of $21^{st}$ century multiple countries around the world, including France, developed programs aiming at the mass deployment of smart meters~\citep{wildSmartMeter2021,ademeCompteurLinkyDeploiement2021}.
By the end of 2021 about 34 million smart meters, called Linky, were installed in France~\citep{franceInstallationLinkySmart2021}.

\paragraph{Signal disaggregation}
Linky records energy consumption data at intervals of 30 and, in some cases 10 minutes, and uploads it to the data storage infrastructure of Enedis, the French energy distribution company, through power-line communication~\citep{horowitzPowerSystemsRelaying2008}, while technical and contractual data is retrieved from energy providers.
Since the characteristic operational time scales of the most appliances are in the order of minutes, disaggregation of the total signal is a difficult task.
In addition to that, publicly available ground truth datasets are scarce and often cover only a few specific appliance types, e.g. only refrigerators, or do not provide sufficient samples of appliance instances within a certain class, e.g. only one type of refrigerators, or few refrigerators at all, limiting the applications of supervised methods.
Moreover, different classes of appliances, such as, for example, electric vehicles or laundry driers, are distributed unevenly across countries.
In France a variety of options are available to individuals for heating water, with most popular modes being cumulative water heaters (CWH), gas heaters and collective heating.

\paragraph{CWHs as targets for disaggregation}
Cumulative Water Heaters are good targets for disaggregation due to the following reasons:
\begin{enumerate}[(a)]
    \item In the residential sector, water heating is one of the main contributors to electrical consumption (about 12\%~\citep{ademeEauChaudeSanitaire2016}).
    \item Its average "on" time is typically greater than the time resolution of low frequency meters, such as Linky.
Typical volume is in the range of 40 to 300 liters, and the average time required to heat a full load of water to operational temperature varies between $1.5$ and $5$ hours depending on volume and power~\citep{trinhHowLongDoes2021}.
    \item Most electrical energy providers for individuals in France offer contracts with split pricing, where the price of energy is about 25 to 30\% cheaper during off-peak hours.
    The off-peak hours are reported from the energy provider to the meter through power-line communication.
Then users have the option to synchronize their use of appliances with the off-peak hours, which can be achieved automatically, either via a timer or a signal from a pilot wire connected to a device that detects the activation of off-peak hours using a signal sent through the electrical mains~\citep{manomanoCommentInstallerContacteur2017}.
\end{enumerate}

We therefore expect a CWH, correctly connected to the pilot wire, to reliably switch on at the beginning of each off-peak interval and remain in that state until the full tank is heated to its target temperature.
Under normal usage each "on" cycle lasts more than 30 minutes which is the time resolution of Linky data~\citep{trinhHowLongDoes2021}.

\medskip Hello Watt is an energy consulting firm for individuals, that also provides monitoring and analysis services.

Individuals may opt-in to share their energy consumption data and some of their metadata, such as the type of water heating, through a consent form signed per household as mandated by French and European laws.
The insights drawn from the analysis of large representative data samples helps to real-world user problems, reduce energy consumption and bills, improve the quality of life.
In this paper we present a novel disaggregation method for CWHs in Section~\ref{sec:disaggregation-method-for-devices-with-regular-time-signatures} and study its performance on a labeled dataset in Section~\ref{sec:results}.
In Section~\ref{sec:the-dataset} we present a large anonymized sub-sample of Hello Watt user consumption data ($\sim$ 5k users) (and make it publicly available) and in Section~\ref{sec:results_hw} the results of application of our CWH model to this dataset.
We conclude with a brief discussion in Section~\ref{sec:discussion-and-perspectives}.

\section{Related Work}

Disaggregation of energy consumption curves is a main component of Non-Intrusive Load Monitoring (NILM), which is an active research area.

When the aggregated load curve is available at a high time resolution, many approaches are possible such as event matching, where device activation edges are identified and matched to appliances\ \citep{aziziEventMatchingClassification2021}.
Deep learning has also been used to successfully disaggregate load curves using training data\ \citep{mauchNewApproachSupervised2015,kimNonintrusiveLoadMonitoring2017,rafiqRegularizedLSTMBased2018,kimApplianceClassificationPower2019}.
These approaches use deep recurrent neural networks such as Long Short-Term Memory, but some solutions also exist that use only feed-forward networks\ \citep{zhangSequencetopointLearningNeural2017,chenConvolutionalSequenceSequence2018,zhangNonintrusiveLoadMonitoring2019}.
Other approaches model the behavior of the building as a Hidden Markov Models where combinations of device activations are modeled as states of a Markov chain which is partially observed\ \citep{makoninRealTimeEmbeddedLowFrequency2014}.
The last approach is especially relevant in our case because it has been successfully applied to sampling rates as low as one sample per minute.

However, our use case requires algorithms that can process even lower sampling rates down to one sample per 30 minutes.
At this resolution, most approaches are device-specific.
For instance, the authors of\ \cite{culiereBayesianModelElectrical2020b} use the statistical relationship between daily energy consumption and exterior temperature to disaggregate the electrical heating component.

\section{Disaggregation Method for Devices with Regular Time Signatures}\label{sec:disaggregation-method-for-devices-with-regular-time-signatures}

Our disaggregation method is based on the observation that CWH consumption signal has regular features: time signature of activations linked to beginning of off-peak hours and sufficiently long active periods.
The method consists of two parts.
First, we compute the threshold for background consumption and tag data points that exceed this threshold as candidate spikes \emph{(spike detection)}.
In the second step we identify whether there exists a pattern of spikes that matches the expected characteristics of the device of interest \emph{(spike filtering)}.

The input of our model is composed of an evenly spaced times series representing average power consumption, a set of monotonic intervals describing off-peak hour ranges $\{(t^{(k)}_a, t^{(k)}_b)\}$, where $ 0 \le t^{(k)}_a < t^{(k)}_b < 24$,
and a range for the expected power level of the device of interest.

The output of our model is a list of activation times, durations and power usage of the device of interest.

In this publication all power levels are expressed in kW\@.
Time series are given with a frequency of 30 minutes, and off-peak range boundaries are rounded to 30-minute increments.

\paragraph{Spike detection}
The first step of our pipeline is to compute a background noise threshold, identify spikes that exceed it and store whether their start coincides with the start of off-peak hours.

\begin{figure}
    \centering
    \includegraphics[width=.8\textwidth]{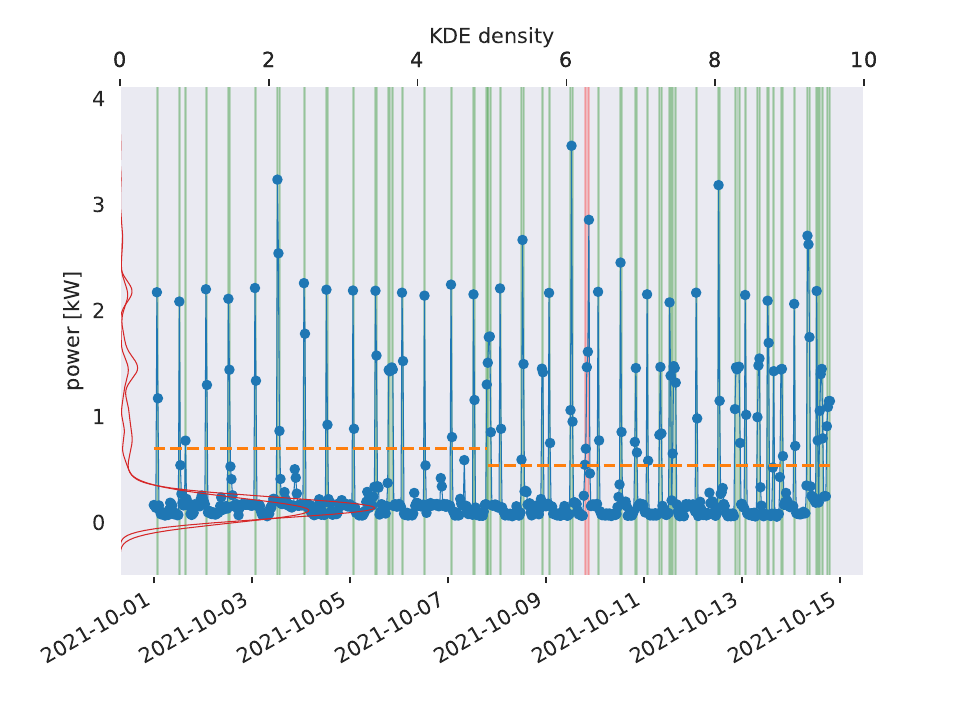}
    \caption{Signal thresholding, spike identification and filtering over two weeks of observations. The load curve (in blue, referenced on the bottom x-axis) is partitioned between base consumption and peak by a threshold (dashed orange line) computed using KDE (red line, referenced on the top x-axis). The background color of a spike indicates whether it was selected in the filtering step of the algorithm (green for valid and red for invalid spikes).}
    \label{fig:signal_kde}
\end{figure}

\begin{enumerate}
    \item We split the time series into a list of smaller sections that are handled independently (in our application we split the input into 7-day segments).
    These sections are later referred to as observations.
    This step reduces the sensitivity of downstream steps to the amount of input data, and handles consumption drift.

    \item We apply a Kernel Density Estimator (KDE) to each observation independently, and use the first local minimum of the estimator as the threshold between background and spike consumption\footnote{\url{https://stackoverflow.com/a/11516590}}.
    We used Scott's rule to select the KDE bandwidth~\citep{scottMultivariateDensityEstimation1992}, and this process is depicted in~\Cref{fig:signal_kde}, where the red curve represents the KDE.

    \item For each observation, we select sequences of consecutive consumption values that are above the threshold.
    These are referred to as spikes.
    For each spike we store the start time of the spike relative to the closest start of off-peak hours as well as the maximum power used during the spike, minus the local background power level. 
    For this last step, we estimate the background power level as the average of two closest data points that are outside the spike.
    Indeed, as is visible in \Cref{fig:signal_kde}, the computed threshold is well above the background level and only serves as a classification boundary.
\end{enumerate}

\begin{figure}
    \centering
    \includegraphics[width=.8\textwidth]{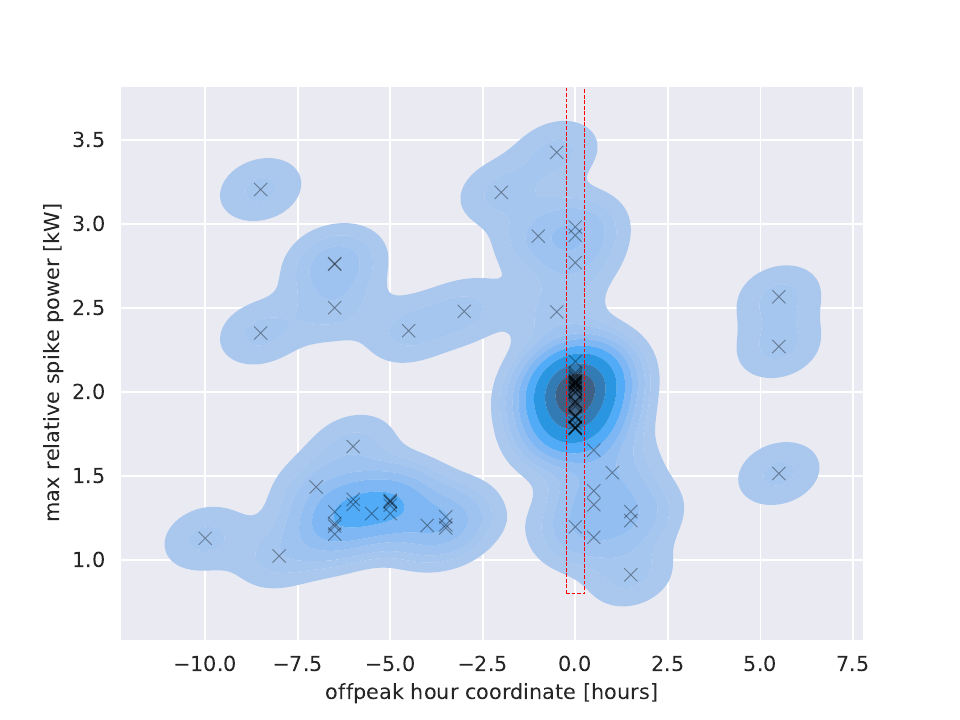}
    \caption{Density plot of power spikes for a single home, with time relative to the nearest off-peak segment start as X-axis, and energy as Y-axis.
    It shows a dense cluster of points at the beginning of off-peak hours and with a maximum power of 2 to 4 kW, consistent with the presence of CWHs connected to an off-peak power switch.}
    \label{fig:energy_offpeak_coord}
\end{figure}

\begin{figure}
    \centering
    \includegraphics[width=.8\textwidth]{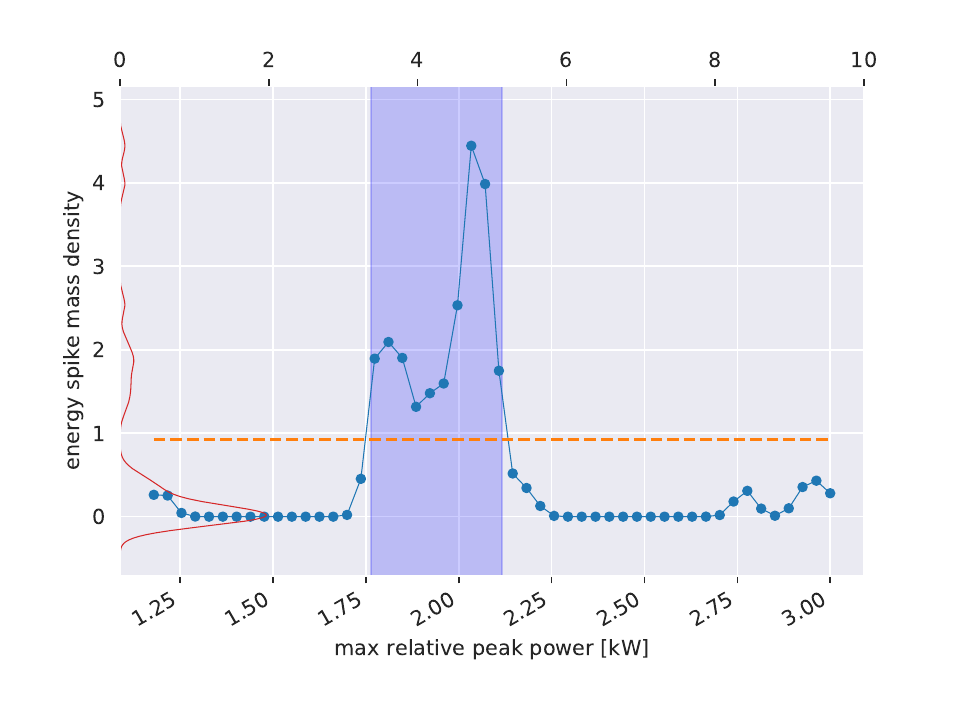}
    \caption{Second thresholding step in order to identify the range of spike powers observed at the start of off-peak hours, using KDE. The occurrence rate of each peak power level is represented as a blue curve, then a KDE is constructed based on the occurrence rates. The KDE distribution is shown in red, and is referenced on the top x-axis. The lowest local minimum of this KDE is then used as a threshold (displayed as a dashed line).}
    \label{fig:energy_kde}
\end{figure}

\paragraph{Spike filtering}
\Cref{fig:energy_offpeak_coord} shows a density plot of the characteristics of the spikes identified in the previous step.
The filtering step of the algorithm identifies whether a cluster matching the expected characteristics of a CWH exists in this space.

Two conditions must be met for a spike to be classified as being caused by a CWH:
(a) it starts exactly at the beginning of off-peak hours (this is represented as a red box in \Cref{fig:energy_offpeak_coord});
(b) the set of all spike power levels that validate criteria (a) form a cluster within the expected range for the target device (in our case 0.8 to 5~kW).

The clustering step in (b) uses KDE again to find a peak in the space of power levels, much like the spike identification step uses KDE to find spikes in time.
The second clustering step is depicted in~\Cref{fig:energy_kde}.

If such a range is found matching the expected power of the device of interest, then the fit is considered successful and all matching power spikes can be attributed to this device.

\section{Model Validation}\label{sec:results}

We equipped a single home with a custom electrical meter that provides a ground truth by measuring the power of the CWH about once per second.
In parallel, the overall consumption of this home is independently reported through its smart meter and collected from the energy provider at a resolution of 30 minutes.

\begin{figure*}
    \centering
    \includegraphics[width=\linewidth]{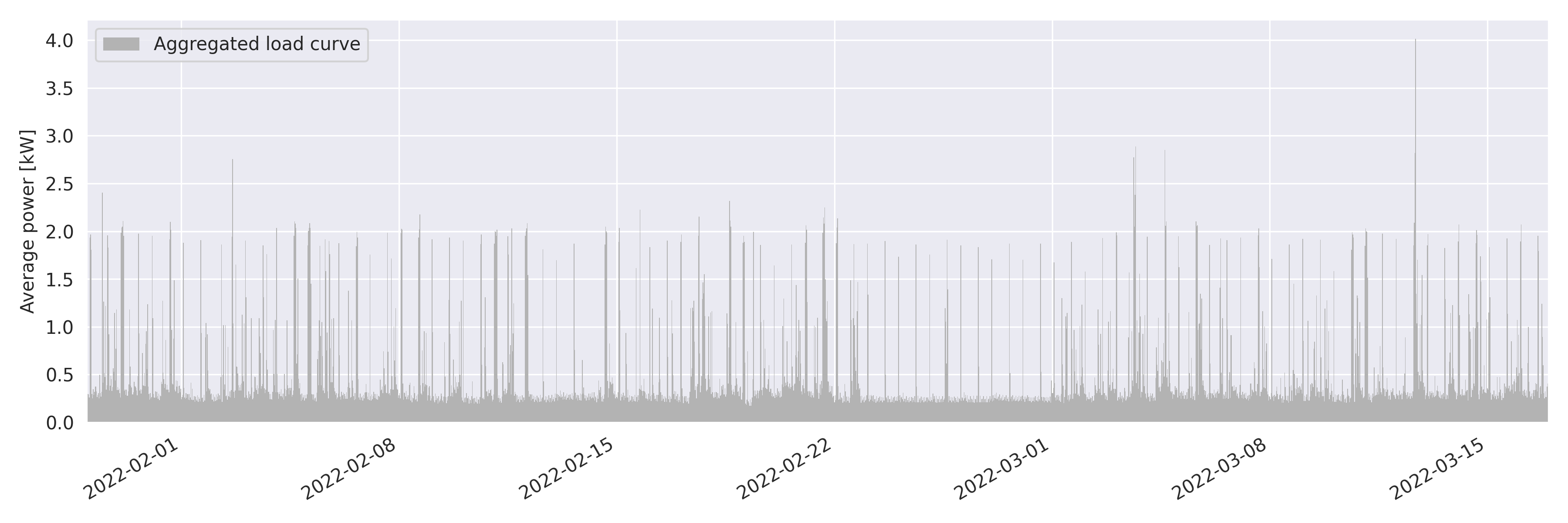}
    \includegraphics[width=\linewidth]{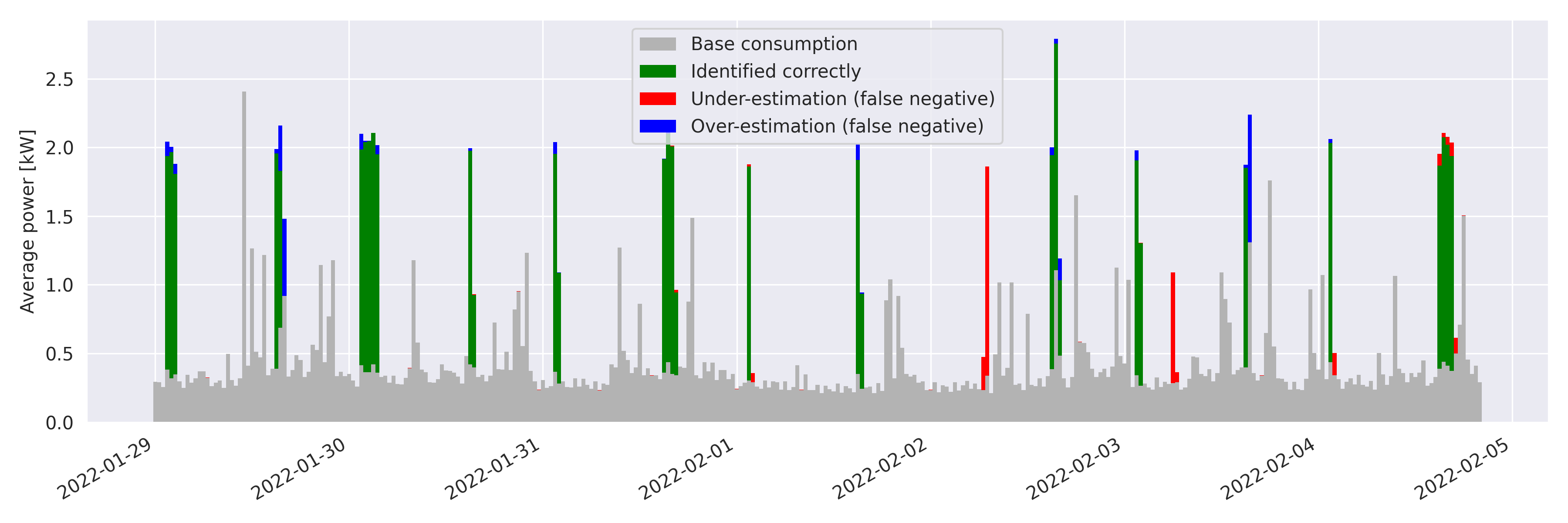}
    \caption{Top: Full load curve of the house for which we have ground-truth water heater consumption data.\\ Bottom: First 7 days of the same load curve. For each 30-minute interval, the portion of consumption that was identified correctly as CWH is green, the portion that was identified correctly as non-CWH is grey, portions that were mistakenly identified as CWH are blue and portions that were mistakenly identified as non-CWH are red.}
    \label{fig:inhouse_full_cdc}
\end{figure*}

Therefore, we were able to test the disaggregation model against a known baseline. \Cref{fig:inhouse_full_cdc} shows the load curve of the home during the testing period, and colors are superimposed to show over and under-estimations of the water heater consumption according to our model.

\begin{figure}
    \centering
    \includegraphics[width=.7\linewidth]{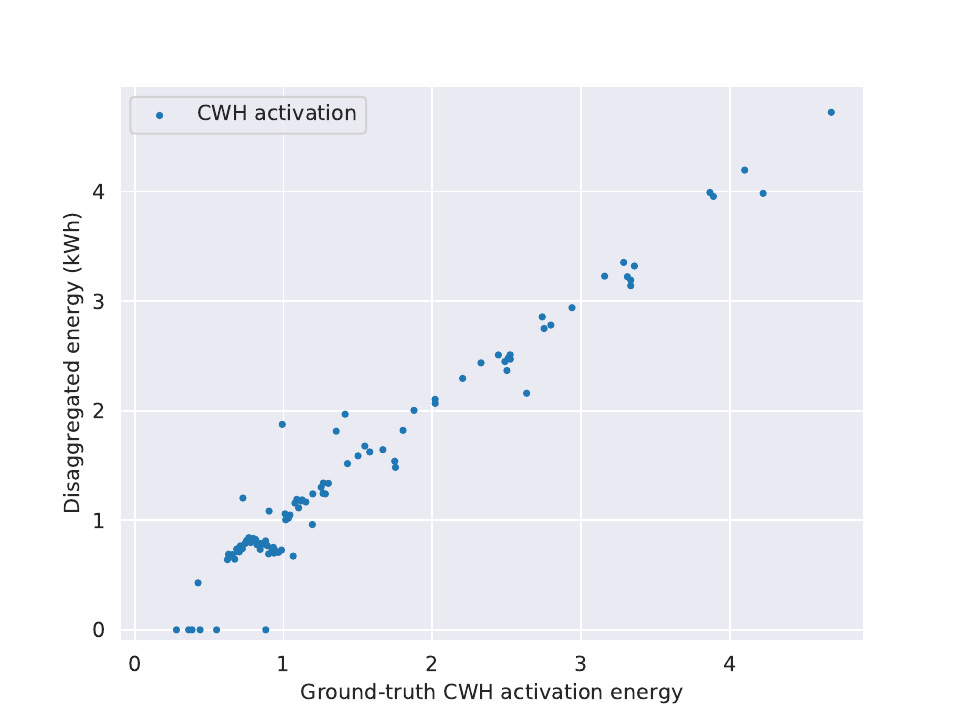}
    \caption{Predicted vs ground-truth power of CWH for each CWH activation (an activation is defined as a series of consecutive 30-minute intervals where the average ground-truth power of the CWH is greater than 100 watts).}
    \label{fig:inhouse_scatter_spikes}
\end{figure}

\Cref{fig:inhouse_scatter_spikes} shows activations of the water heater (an activation being defined as a series of consecutive 30-minute intervals where the average power of the CWH is greater than 100 watts).
It demonstrates that most of the activations of the water heater were properly detected, the only exceptions being very short activations that occur early in the morning when the water heater re-activates before the end of the off-peak period (for instance when someone takes an early shower).
No false positives were identified in terms of activations (all detected activations correspond to a ground-truth one), meaning that when viewed as a test for activations of the CWH, our model has a precision of 100~\%.
The false negatives amount to a recall of \input{figs_generated/activations_recall.tex}.

\begin{figure}
    \centering
    \includegraphics[width=.7\linewidth]{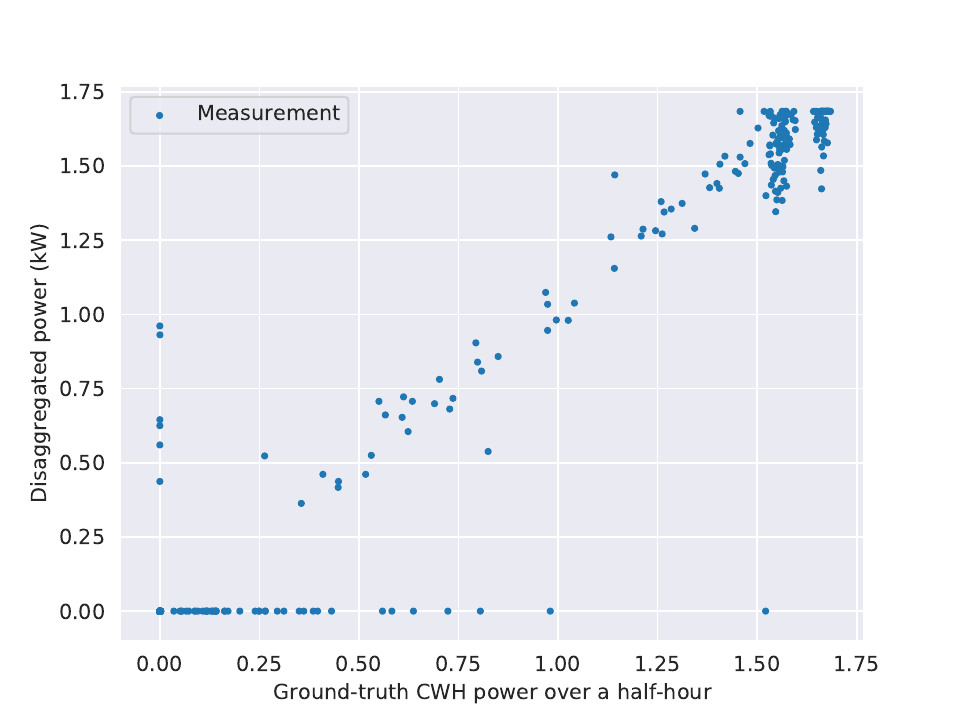}
    \caption{Predicted vs ground-truth power of CWH for each half-hour period.}
    \label{fig:inhouse_scatter}
\end{figure}

Conversely, \Cref{fig:inhouse_scatter} shows each individual 30-minute period, and we can see both some false positives where CWH consumption was wrongly identified, and some false negatives where an interval was wrongly identified as not being part of a CWH activation.

For a given 30-minute interval, we use our model to detect whether the CWH used more than 100 watts on average.
The statistics of this test are presented in \cref{tab:table_spikes}.

However, these over and under-estimations happen mostly at the trailing edge of CWH activations: the activation itself is recognized with high accuracy, but its predicted duration is not always exact, especially when it ends early in the last half-hour period.
In these cases the last 30-minute interval blends into the background and is hard to identify, however this has little effect on the computed energy of the activation as a whole.

\begin{table}
    \centering
    \input{figs_generated/table_spikes}
    \caption{Ability of our model to predict for a single 30-minute interval whether the average power level used by the CWH is above 100 watts.}
    \label{tab:table_spikes}
\end{table}

\section{Hello Watt Dataset Description}\label{sec:the-dataset}

Hello Watt collects power usage data at a resolution of 30 minutes.
Among all users approximately 46\% of contracts are on the off-peak pricing.
Among only users who declared using an electrical water heater, this fraction reaches 66\%.
The distribution of water heating types, and the fraction of contracts with off-peak pricing is presented in \Cref{tab:wheating_type}.

\begin{figure}
    \centering
    \includegraphics[width=.4\textwidth]{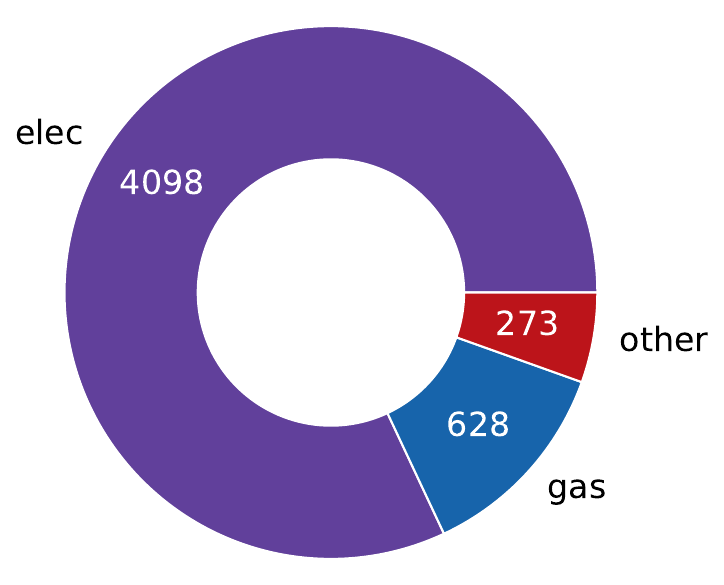}
    \caption{Water heating types of houses selected for analysis.}
    \label{fig:wheating_type_sub}
\end{figure}

To develop and test our disaggregation method we consider a subsample consisting of power consumption of 5k households with off-peak pricing contracts for one month.
The number of households per water heating type in this dataset is presented in \Cref{fig:wheating_type_sub}.
This dataset is published through the Open Science Framework\footnote{\url{https://osf.io/ebjd3/?view_only=54c6097e95984b6c9c933a38e574a8f7}}.
We chose to include in our dataset both homes with and without electrical water heaters to validate the results of the model as presented in~\Cref{sec:results}.

In addition to the type of their water heating, some users also provide such metadata as the home surface area, and the number of inhabitants.

Our working subsample consists of electricity consumers interested in their energy consumption, and therefore it may have some have a distributional bias.

\begin{table}
    \centering
    \input{figs_generated/table_whtype_hphc}
    \caption{Water heating and pricing types in the full Hello Watt database. HPHC pricing is short for "Heures Pleines / Heures Creuses" which means separate pricing for on-peak and off-peak hours.}
    \label{tab:wheating_type}
\end{table}

\begin{figure}
    \centering
    \includegraphics[scale=0.5]{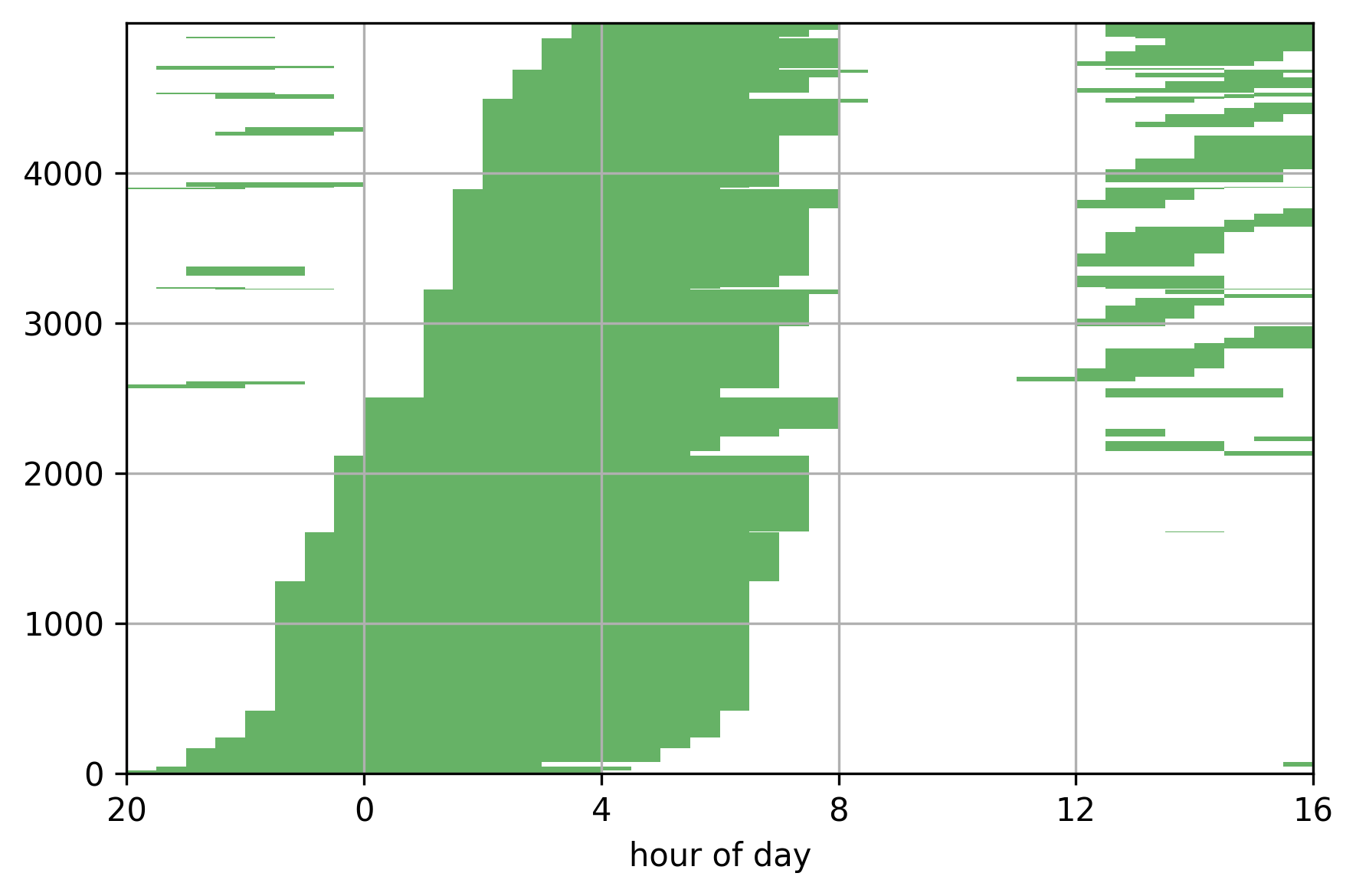}
    \caption{The distribution of off-peak ranges for the Hello Watt subsample.}
    \label{fig:hc_dist}
\end{figure}

\paragraph{Peak / off-peak contracts}
For a given house and day, off-peak hours are made of up to three intervals, whose total duration is exactly eight hours.
These times are always expressed in local time (timezone Europe/Paris) and for a given house they are usually identical for all days.
The distribution of off-peak ranges in our sub-sample is presented in \Cref{fig:hc_dist}.
The off-peak hours are split in two intervals in 92\% of houses, one continuous interval for 4\% of houses, and three intervals for the remaining 4\% .

\section{Analysis of Hello Watt Dataset}\label{sec:results_hw}

In this section, we run the model on a subset of houses from our database.
These consist of a random sample of houses that meet the following criteria:

\begin{itemize}
    \item Have at least 1440 ($48\times 30$) energy measures during the month of october 2021 (october has 31 days so this allows up to 3\% of missing data).
    \item The technical and contractual data we retrieve from the energy provider indicates that the user is billed differently for off-peak hours.
    \item The off-peak hours are known for this house.
\end{itemize}

Each home can have one or several off-peak hour ranges, either during the night or the afternoon. \Cref{fig:hc_dist} shows all the off-peak ranges of the houses in the dataset.
For each home, we also have self-reported data about the main mode of water heating, surface and number of inhabitants.
\Cref{fig:wheating_type_sub} shows self-reported water heating modes in the dataset.

This allows us to test the ability of our model to distinguish homes with and without CWH by comparing its output with self-reported data.

\subsection{CWH Detection}\label{subsec:cwh-detection}

\begin{figure}
    \centering
    \includegraphics[width=.4\linewidth]{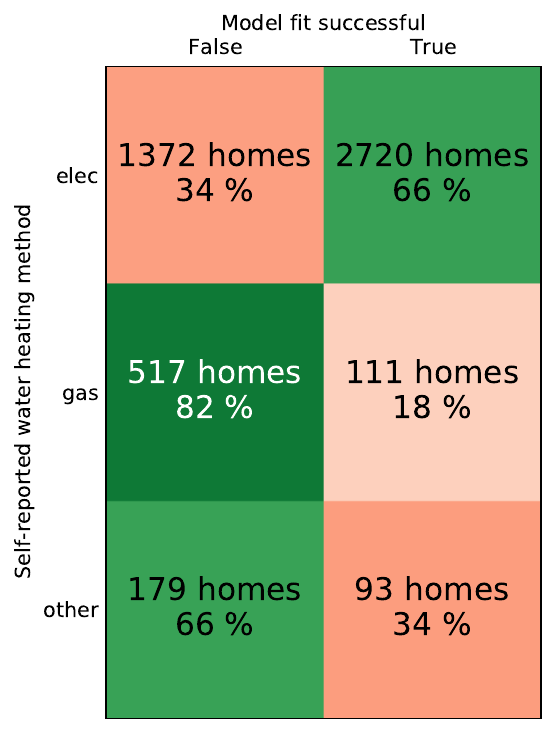}
    \caption{Fraction of homes in which a CWH was found depending on self-reported water heating types.}
    \label{fig:wh_predict_vs_declared}
\end{figure}

The results presented in \Cref{fig:wh_predict_vs_declared} reveal that in \input{figs_generated/wh_pvd_tp_pct} of cases where the user has declared having an electrical water heating we detect a device with on/off cycles compatible with off-peak hour triggering hypothesis.
In houses where the user declared water heating is achieved through gas, this fraction drops to \input{figs_generated/wh_pvd_fp_pct}.
The differences between the self-reported data and the model results can be attributed to both model imperfection and misreporting by users of their devices.

\begin{figure}
    \centering
    \includegraphics[scale=0.5]{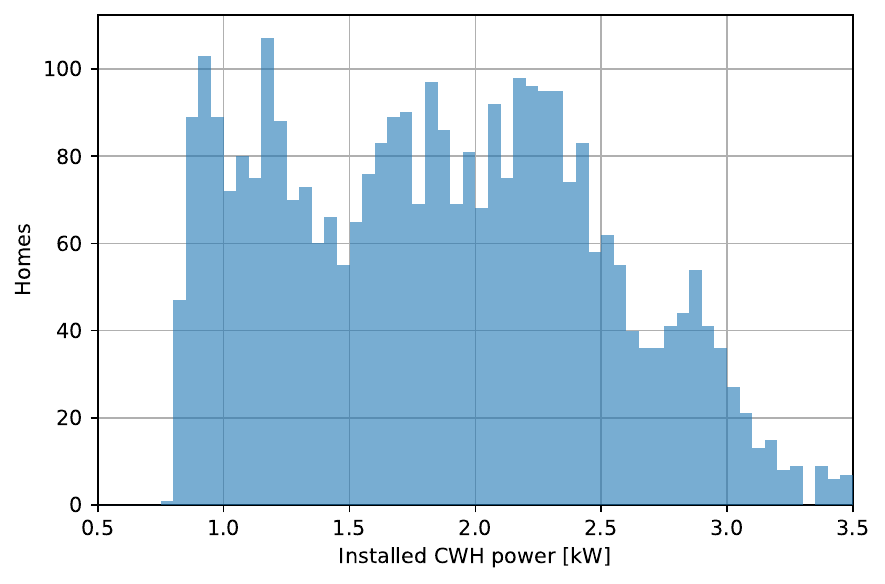}
    \caption{Power levels of CWHs identified by our model.}
    \label{fig:dist_power}
\end{figure}

In \Cref{fig:dist_power} we show the power levels of identified devices.
Several peaks are visible in this graph despite the noise, and CWH models can be classified as low power (.6 to 1.5~kW), medium power (1.5 to 2.7~kW) and high power (2.7 to 3.3~kW).

\subsection{CWH Disaggregation Results}
\label{subsec:consumption-disaggregation-results}

\begin{figure}
    \centering
    \includegraphics[scale=0.5]{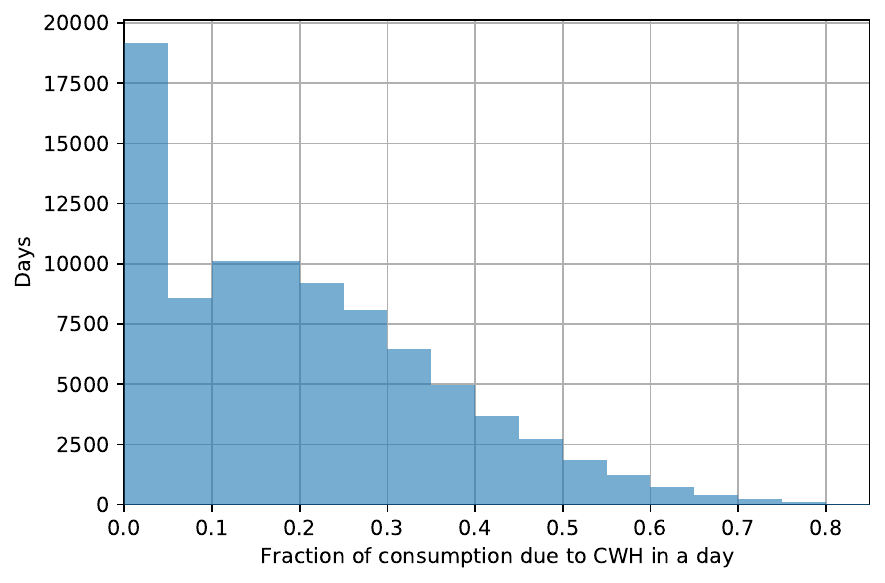}
    \caption{Distribution of daily consumption of CWHs identified by our model, as a fraction of the total consumption for that day.}
    \label{fig:dist_daily_average}
\end{figure}

\begin{figure}
    \centering
    \includegraphics[scale=0.5]{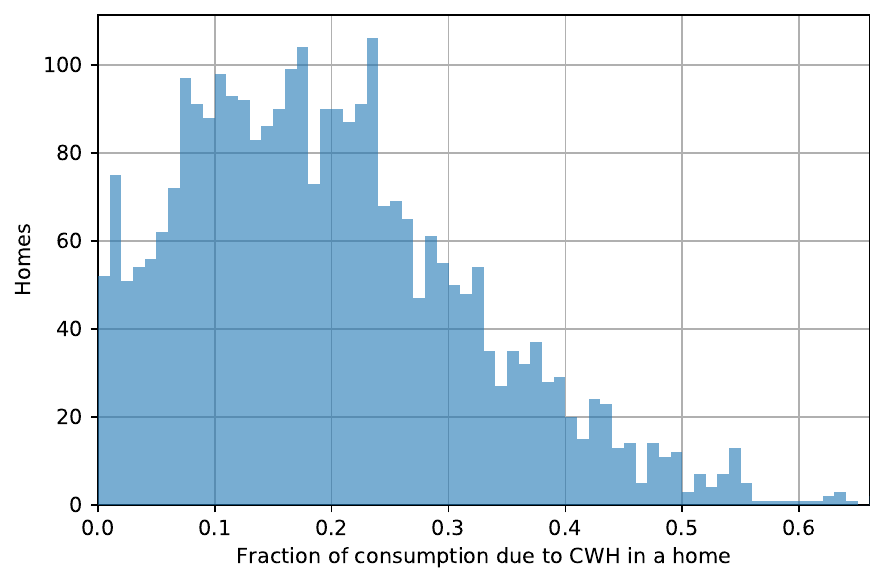}
    \caption{Distribution of total consumption of CWHs identified by our model, as a fraction of the total consumption of the household.}
    \label{fig:dist_overall_average}
\end{figure}

In \Cref{fig:dist_daily_average,fig:dist_overall_average}, for each household in our dataset we estimate the daily and overall consumption fraction due to identified CWHs.

The spike at zero in the distribution of daily average corresponds to days when there was no hot water usage.
It is possible that tenants are away a fraction of time, we also note that our current model does not discover all the relevant peaks, especially if local background values are excessive.

The most prominent spike in the overall consumption plot in the range 8\%-14\%, in agreement with the estimated 15\% provided by the energy agency ADEME~\citep{ademeEauChaudeSanitaire2016} and the range between $12\%$ and $15\%$ given by France's historical energy provider EDF~\citep{edfConsommationGazUsages}\@.

\section{Discussion and Perspectives}
\label{sec:discussion-and-perspectives}

In this study we presented a simple model, based on domain knowledge, that disaggregates signals that have quasi-regular time signatures and applied it to the dataset that purportedly contains a large fraction of such appliances, namely CWHs. 

Overall we find consistent results, with CWH power ranging from 1 to 3.5~kW\@.
The disaggregated fraction of energy used by CWHs is consistent with available estimates.
Our model correctly identifies CWHs predominantly where we expect them, i.e.\ in cases where users declare electrical water heating.

The performance of our model is validated on a ground-truth dataset and is deemed adequate with precision and recall $>0.9$.
Although it has been tested for time series with 30 minutes resolution, we expect our model to work in wide range of resolutions from one minute to one hour.

We note the following shortcomings of our approach:
\begin{itemize}
    \item The CWH model is based on spike identification with respect to local background consumption and assumes the CWH activations are far more frequent at the beginning of off-peak hours compared to activations of other devices.
        In cases where several devices are systematically activated at the beginning of off-peak hours, (the second one could be a washing machine for instance) our model may fail at separating the CWH, if coupled activations are more prevalent.
    \item Our model works under the assumption that there is only a single off-peak hour triggered device that has only one mode.
        Multiple spikes may correspond to the modes of the same CWH or to activations of several devices.
    \item While the performance metrics of our model with respect to ground truth data are very promising, our testing dataset is limited to the consumption load curve of a single household.
    \item Our model is aimed at detection of CWH peaks for homes that have split off-peak hours pricing.
        However, nearly one third of owners of electrical water heaters do not use such contracts.
\end{itemize}

\section{Conclusion}\label{sec:conclusion}

In this project we developed an unsupervised model of CWH disaggregation and validated its performance on a labeled dataset.
We compiled and made public a large dataset of power consumption of $\sim$ 5k households relevant for CWH studies.
Our model was applied to this dataset to characterize CWH consumption at scale in a non-intrusive manner.
This simple non-intrusive load monitoring model enables identification and characterization of CWHs performance.

On the one hand, it may be used to identify incorrectly installed of CWHs in regard to peak/off-peak contracts:
this is the case when a user is on a peak/off-peak hour contract, but our model does not identify a regular consumption signature due to a declared CWH.
Such misconfiguration has a direct quantifiable economic impact.
On the other hand, our model can be used to track the consumption of a given user and monitor gradual or abrupt changes in the water heating energy consumption and prompt maintenance recommendations.

Large scale analysis of the consumption due to CWHs made possible thanks to our model allows the characterization of the overall power distribution of electrical water heaters as well as consumption patterns and may be used for further policy proposals in the view of governmental energy saving programs.

\section{Acknowledgements}

We thank our colleagues Laetitia Leduc and Xavier Coudert for fruitful discussions and support during this project.
The energy consumption data of participating households was provided by our partner Enedis.

\bibliographystyle{elsarticle-num}
\bibliography{wh}

\newpage
\appendix

\section{Dataset Anonymization}

The dataset we published is composed of consumption data at a 30-minutes resolution for 5k homes.
It also includes metadata such as the number of inhabitants, surface, water heating type and off-peak hours for the household.

Our dataset was anonymized with the following goals in mind:
\begin{itemize}
  \item Prevent a single individual from being identified through this data.
  \item Make re-identification non-trivial even if someone already holds part of the published data from another source.
  \item Adhere to the principle of data minimization and publish only data that has scientific interest.
\end{itemize}

The sample was limited to one month of data, any internal identifiers were replaced with random strings, and a small random noise was added to each power value.
This noise follows an exponential distribution with an expected value of 0.005~kW\@.

\end{document}

%% file: figs_generated/activations_recall.tex
93.9~\%

%% file: figs_generated/table_spikes.tex
\begin{tabular}{lr}
\toprule
True positives  &     197 \\
True negatives  &    2022 \\
False positives &       6 \\
False negatives &      35 \\
Precision       &  97.0 \% \\
Recall          &  84.9 \% \\
\bottomrule
\end{tabular}

%% file: figs_generated/table_whtype_hphc.tex
\begin{tabular}{lrrrr}
\toprule
{} &   base &   hphc & hphc frac. &         Total \\
\midrule
elec  &   7471 &  14573 &       66 \% &  22044 (57 \%) \\
gas   &  10625 &   1960 &       16 \% &  12585 (33 \%) \\
other &   2776 &   1007 &       27 \% &   3783 (10 \%) \\
Total &  20872 &  17540 &       46 \% &         38412 \\
\bottomrule
\end{tabular}

%% file: figs_generated/wh_pvd_tp_pct.tex
66.5 \%

%% file: figs_generated/wh_pvd_fp_pct.tex
17.7 \%